\documentstyle[11pt,dina4]{article}

\newcommand{\eqs}[2]{Eqs.~(\ref{#1})--(\ref{#2})}
\newcommand{\fig}[1]{Fig.~\ref{#1}}
\newcommand{\tab}[1]{Table \ref{#1}}
\newcommand{\subscript}[2]{#1_{\mbox{\scriptsize{#2}}}}
\newcommand{\D}{\displaystyle}

\newcommand{\tr}{\mbox{tr}\,}
\newcommand{\diag}{\mbox{diag}\,}

\newcommand{\lagr}{{\cal L}}

\newcommand{\pp}{\pi^{+}}
\newcommand{\pmi}{\pi^{-}}
\newcommand{\pn}{\pi^{0}}
\newcommand{\ppm}{\pi^{\pm}}
\newcommand{\pmp}{\pi^{\mp}}
\newcommand{\kp}{K^{+}}
\newcommand{\km}{K^{-}}
\newcommand{\kn}{K^{0}}
\newcommand{\knb}{\:\overline{\! K^{0} \!}\:}
\newcommand{\en}{\eta}

\newcommand{\massek}{M_{K}}

\newcommand{\klong}{\subscript{K}{L}^{0}}

\newcommand{\mup}{m_{u}}
\newcommand{\md}{m_{d}}
\newcommand{\ms}{m_{s}}

\newcommand{\uq}{u}
\newcommand{\saq}{\overline{s}}

\newcommand{\kla}{\big<}
\newcommand{\mer}{\big>}
\newcommand{\sta}{\big|}

\input{psfig}

\begin{document}

\hfill hep-ph/9410368\\
\vspace{0.5cm}
\begin{center}
{\LARGE $K_{e5}$ decays  in chiral perturbation theory} \\
\vspace{0.7cm}
Stefan Blaser \\
Institut f\"ur theoretische Physik,
Universit\"at Bern\\
Sidlerstrasse 5, CH-3012 Bern\\
blaser@butp.unibe.ch\\
\vspace{0.5cm}
October 1994
\vspace{0.7cm}
\end{center}

\begin{abstract}
We evaluate the branching ratios for the decays $K\rightarrow \pi\pi\pi e\nu$
at leading order in chiral perturbation theory and give an isospin relation
for the decay rates.
\end{abstract}

 {\bf 1}. We discuss the $K_{e5}$ decays
\begin{eqnarray*}
    \kp &\longrightarrow& \pp \pmi \pn e^{+} \nu_{e} ,
\\
    \kp &\longrightarrow& \pn \pn \pn e^{+} \nu_{e} ,
\\
    \kn &\longrightarrow& \pn \pn \pmi e^{+} \nu_{e} ,
\\
    \kn &\longrightarrow& \pp \pmi \pmi e^{+} \nu_{e}
\end{eqnarray*}
in the framework of chiral perturbation theory (CHPT)
\cite{weinberg,gasser84}.
For low momenta relevant in the present case, the
transition amplitude for
$ K \rightarrow  \pi \pi \pi e^{+} \nu_{e} $
reduces in the standard model to the current times current form
\begin{equation}                                                 \label{gl1}
      T = \frac{\subscript{G}{F}}{\sqrt{2}} V^{\ast}_{us}
        \overline{u}(p_{\nu}) \gamma_{\mu} (1-\gamma_{5})
        v(p_{e}) (V^{\mu}-A^{\mu}),
\end{equation}
where
\begin{equation}                                                  \label{gl2}
    V^{\mu}-A^{\mu} = \kla \pi(p_{1}) \pi(p_{2}) \pi(p_{3}) \mbox{ out}
              \sta \saq \gamma^{\mu} (1-\gamma_{5}) u
              \sta K(p) \mer.
\end{equation}

{\bf 2}. To calculate the hadronic matrix elements $V^\mu$ and $A^\mu$,
we use the effective Lagrangian of QCD at leading order,
\begin{equation}                                                  \label{gl4}
  \lagr = \frac{F^{2}}{4} \: \tr \left(
                             \partial_{\mu}U \partial^{\mu}U^{\dagger}
                             + \chi U^{\dagger} + \chi^{\dagger} U
                                  \right),
\end{equation}
where $F=93.2$ MeV is the pion decay constant in the chiral limit.
Furthermore, we work in the isospin limit, {\it i.e.},
$\mup = \md \doteq \hat{m}$, and set
$\chi = 2 B_{0} \, \diag(\hat{m},\hat{m},\ms)$,
where $B_{0}$ is related to the quark condensate in the chiral limit
\cite{gasser84}.
The unitary $3\times3$ matrix $U$ incorporates the fields of the eight
pseudoscalar Goldstone bosons. A convenient parametrization is
$U = \exp(i \sqrt{2} \Phi /F)$ with
\begin{equation}                                                  \label{gl5}
      \Phi =            \left(
                        \begin{array}{ccc}
      \D \frac{\pn}{\sqrt{2}} + \frac{\en}{\sqrt{6}} & \D \pp & \D \kp \\
      \D \pmi & \D -\frac{\pn}{\sqrt{2}} + \frac{\en}{\sqrt{6}}& \D  \kn \\
      \D \km & \D \knb & \D -\frac{2}{\sqrt{6}} \en
                        \end{array}
                      \right).
\end{equation}
In addition, the vector current relevant in the present case reads
\begin{equation}
      V_{\mu}^{4-i5} =  \frac{F^{2}}{4i} \;
                       \tr \left( [\lambda^{4}-i\lambda^{5}]
                                  [ U\partial_{\mu}U^{\dagger}
                                       +U^{\dagger}\partial_{\mu}U]
                           \right),
\end{equation}
where $\lambda^{4}$ and $\lambda^{5}$ denote
Gell-Mann matrices. The corresponding axial vector
current does not contribute, because it is odd under the transformation
$\Phi\rightarrow -\Phi$.

 {\bf 3.} The relevant Feynman diagrams at leading order  in CHPT are shown in
\fig{figur1}.
\begin{figure}
\begin{center}
\setlength{\unitlength}{0.0125in}%
\begin{picture}(422,117)(119,512)
\put(119,512){\strut\psfig{figure=graph.ps}}
\put(316,575){\makebox(0,0)[lb]{\raisebox{0pt}[0pt][0pt]{\elvrm $\pi$}}}
\put(481,592){\makebox(0,0)[lb]{\raisebox{0pt}[0pt][0pt]{\elvrm $K$}}}
\put(431,595){\makebox(0,0)[lb]{\raisebox{0pt}[0pt][0pt]{\elvrm $V^{\mu}$}}}
\put(270,595){\makebox(0,0)[lb]{\raisebox{0pt}[0pt][0pt]{\elvrm $V^{\mu}$}}}
\put(541,558){\makebox(0,0)[lb]{\raisebox{0pt}[0pt][0pt]{\elvrm $\pi$}}}
\put(532,614){\makebox(0,0)[lb]{\raisebox{0pt}[0pt][0pt]{\elvrm $\pi$}}}
\put(480,621){\makebox(0,0)[lb]{\raisebox{0pt}[0pt][0pt]{\elvrm $\pi$}}}
\put(461,539){\makebox(0,0)[lb]{\raisebox{0pt}[0pt][0pt]{\elvrm $K$}}}
\put(384,580){\makebox(0,0)[lb]{\raisebox{0pt}[0pt][0pt]{\elvrm $\pi$}}}
\put(256,539){\makebox(0,0)[lb]{\raisebox{0pt}[0pt][0pt]{\elvrm $K$}}}
\put(310,512){\makebox(0,0)[lb]{\raisebox{0pt}[0pt][0pt]{\elvrm (b)}}}
\put(378,545){\makebox(0,0)[lb]{\raisebox{0pt}[0pt][0pt]{\elvrm $\pi$}}}
\put(366,617){\makebox(0,0)[lb]{\raisebox{0pt}[0pt][0pt]{\elvrm $\pi$}}}
\put(474,512){\makebox(0,0)[lb]{\raisebox{0pt}[0pt][0pt]{\elvrm (c)}}}
\put(165,512){\makebox(0,0)[lb]{\raisebox{0pt}[0pt][0pt]{\elvrm (a)}}}
\put(199,553){\makebox(0,0)[lb]{\raisebox{0pt}[0pt][0pt]{\elvrm $\pi$}}}
\put(214,585){\makebox(0,0)[lb]{\raisebox{0pt}[0pt][0pt]{\elvrm $\pi$}}}
\put(192,613){\makebox(0,0)[lb]{\raisebox{0pt}[0pt][0pt]{\elvrm $\pi$}}}
\put(119,539){\makebox(0,0)[lb]{\raisebox{0pt}[0pt][0pt]{\elvrm $K$}}}
\put(130,595){\makebox(0,0)[lb]{\raisebox{0pt}[0pt][0pt]{\elvrm $V^{\mu}$}}}
\end{picture}
\end{center}
  \caption{\label{figur1}Tree level diagrams for $K_{e5}$ decays.}
\end{figure}
 Their contribution is
\begin{eqnarray}
    \kla \pi^{+}(p_{1}) \pi^{-}(p_{2}) \pi^{0}(p_{3})\mbox{ out}
    \sta \saq \gamma_{\mu} \uq \sta K^{+}(p) \mer
&  =  &
    - A_\mu(1,2,3) + B_\mu(1,2,3),                              \label{gl7} \\
    \kla \pi^{0}(p_{1}) \pi^{0}(p_{2}) \pi^{0}(p_{3})\mbox{ out}
    \sta \saq \gamma_{\mu} \uq \sta K^{+}(p) \mer
&  =  &
      A_\mu(1,2,3)    +   A_\mu(1,3,2)
  +   A_\mu(3,2,1),
                                                                \label{gl8} \\
    \kla \pi^{0}(p_{1}) \pi^{-}(p_{2}) \pi^{0}(p_{3}) \mbox{ out}
    \sta \saq \gamma_{\mu} \uq \sta K^{0}(p) \mer
&  =  &
    \sqrt{2} A_\mu(1,3,2),
                                                                \label{gl9} \\
    \kla \pi^{+}(p_{1}) \pi^{-}(p_{2}) \pi^{-}(p_{3})  \mbox{ out}
    \sta \saq \gamma_{\mu} \uq \sta K^{0}(p) \mer
&  =  &
   -\sqrt{2} \left\{ A_\mu(1,2,3)+A_\mu(1,3,2) \right\},
                                                                \label{gl10}
\end{eqnarray}
where
 \begin{eqnarray}
      A_\mu(1,2,3) & = & \frac{\sqrt{2}}{4 F^{2}}
      \Bigg\{
             \frac{p(p_{3}-p_{1})}{M_{K}^{2} - (p_{1}+p_{3}-p)^{2}}
            [p_{1}-p_{2}+p_{3}-p]_{\mu}  \nonumber \\ &&
\rule{0mm}{8mm}
         {} +  \frac{p(p_{2}-p_{3})}{M_{K}^{2} - (p_{2}+p_{3}-p)^{2}}
            [p_{1}-p_{2}-p_{3}+p]_{\mu}  \nonumber \\ &&
\rule{0mm}{8mm}
         {} -  \frac{p(p_{1}+p_{2})}{M_{K}^{2} - (p_{1}+p_{2}-p)^{2}}
            [p_{1}+p_{2}-p_{3}-p]_{\mu}  \nonumber \\ && 
\rule{0mm}{8mm}
         {} +  \frac{2(M_{\pi}^{2}+2p_{1}p_{2})}
                         {M_{\pi}^{2} - (p_{1}+p_{2}+p_{3})^{2}}
            [p_{1}+p_{2}+p_{3}+p]_{\mu}   \nonumber\\ &&
         {} + 2 [p_{1}+p_{2}-p_{3}]_{\mu}
      \Bigg\}
,                                                               \label{gl11}
 \end{eqnarray}
 \begin{eqnarray}
     B_\mu(1,2,3) & = &  \frac{\sqrt{2}}{4  F^{2}}
      \Bigg\{
             \frac{p(p_{2}-p_{1})}{ M_{K}^{2} - (p_{1}+p_{2}-p)^{2}}
            [p_{1}+p_{2}-p_{3}-p]_{\mu}  \nonumber \\ &&
\rule{0mm}{8mm}
         {} +  \frac{p(p_{2}-p_{3})}{  M_{K}^{2} -(p_{2}+p_{3}-p)^{2}}
            [p_{1}-p_{2}-p_{3}+p]_{\mu}  \nonumber \\ &&
\rule{0mm}{8mm}
         {} +  \frac{p(p_{1}-p_{3})}{ M_{K}^{2} - (p_{1}+p_{3}-p)^{2}}
            [p_{1}-p_{2}+p_{3}-p]_{\mu}
      \Bigg\}
.                                                                \label{gl12}
 \end{eqnarray}

 {\bf 4.} Defining the Lorentz invariant measure
\begin{equation}                                                 \label{gl13}
 \subscript{d}{LIPS}(p;p_{1},p_{2},\ldots,p_{n}) \doteq
 \delta^{4}(p-\sum_{i=1}^{n}p_{i})
 \prod_{i=1}^{n} \frac{d^{3} {\bf p}_{i}}{2 p_{i}^{0}},
\end{equation}
the differential decay  rate is given by
\begin{equation}                                                 \label{gl14}
   d \Gamma = \frac{1}{2 \massek (2 \pi)^{11}}
   \sum_{\mbox{\scriptsize{spins}}} |T|^{2}
   \subscript{d}{LIPS}(p;p_{e},p_{\nu},p_{1},p_{2},p_{3}).
\end{equation}
The rates and branching ratios which follow from \eqs{gl7}{gl14}
are displayed in \tab{tafel1}.
\begin{table}[b] \caption{\label{tafel1}Rates and branching ratios of
$K_{e5}$ decays, evaluated from the leading order term in CHPT.}
\begin{center}
\begin{tabular}{|c|c|c|} \hline
               & decay rate in s$^{-1}$ & branching ratio      \\ \hline
  $\kp \to \pp \pmi \pn  e^{+} \nu_{e}$
               & $2.4 \cdot 10^{-4} $   & $3.0 \cdot  10^{-12}$ \\ \hline
  $\kp \to \pn \pn  \pn e^{+} \nu_{e}$
               & $2.0 \cdot 10^{-4} $   & $2.5 \cdot  10^{-12}$ \\ \hline
  $\klong \to \pn \pn \pmp e^{\pm} \nu_{e}$
               & $2.4 \cdot 10^{-4} $   & $12 \cdot  10^{-12}$ \\ \hline
  $\klong \to \ppm \pmp \pmp e^{\pm} \nu_{e}$
               & $6.5 \cdot 10^{-4} $   & $33 \cdot   10^{-12}$ \\ \hline
\end{tabular}
\end{center}
\end{table}
The smallness of the decay rates is due to the suppression
of phase space. Indeed, consider the ratio of the four- and
five-dimensional phase space volumes
   \begin{equation}
       \frac{\massek^{2}
          \int \subscript{d}{LIPS}
          (p;p_{e},p_{\nu},p_{1},p_{2})}{2! \, (2 \pi)^{12}}
       \times
       \frac{3! \, (2 \pi)^{15}}{\int \subscript{d}{LIPS}
              (p;p_{e},p_{\nu},p_{1},p_{2},p_{3})}
       \approx  2.3 \cdot 10^{6},
   \end{equation}
where we have inserted $\massek^{2}$ for dimensional reasons. On the other
hand, we find for the ratio of the corresponding rates
at tree level in CHPT
   \begin{equation}
     \frac{\subscript{\Gamma(\kp \to \pn \pn e^{+} \nu_{e})}{tree}}%
{\subscript{\Gamma(\kp \to \pn \pn \pn e^{+} \nu_{e})}{tree}}
     \approx 3.4 \cdot 10^{6},
   \end{equation}
which is of the same order of magnitude.

 {\bf 5}. Turning now to the corrections at next-to-leading order, we note
that the matrix element of the axial current receives
a contribution from the chiral anomaly \cite{wess,witten}.
Besides the local term of the Wess-Zumino-Witten action,
also the nonlocal part, which contains at least five meson fields,
gives a contribution. However, an explicit calculation shows
that it is suppressed by the factor $m_{e}$ in the matrix element
and therefore undetectable in the near future.
We expect from experience with other calculations in CHPT that
the remaining contributions to the matrix element (\ref{gl2}) at this
order enhance the tree-level results for the decay rates
at most by a factor of two to three.

 {\bf 6}. Due to isospin symmetry, the relations which are given in
\eqs{gl7}{gl10} are valid to all order in CHPT,
with $A_\mu(1,2,3)$ symmetric in $p_{1}$ and $p_{2}$, and $B_\mu(1,2,3)$
totally antisymmetric in $p_{1}$, $p_{2}$, and $p_{3}$.
The matrix elements of the axial current
$\overline{s}\gamma_\mu\gamma_5u$ have an analogous decomposition,
with the same symmetry properties
of the reduced matrix elements. From this follows the isospin relation
\begin{equation}                                                 \label{gl15}
 2 \, \Gamma(\kp \to \pn \pn  \pn e^{+} \nu_{e})
  =
 \Gamma(\klong \to \ppm \pmp \pmp e^{\pm} \nu_{e})
  -
  \Gamma(\klong \to \pn \pn \pmp e^{\pm} \nu_{e}).
\end{equation}

 {\bf 7}. Hitherto, only poor experimental data on $K_{e5}$ decays are
available. The Particle Data Group \cite{pdg} quotes the upper bound
\begin{equation}
  \frac{\Gamma(\kp \to \pn \pn \pn e^{+} \nu_{e})}{\subscript{\Gamma}{total}}
  < 3.5 \cdot 10^{-6},
\end{equation}
which is six orders of magnitude bigger than our result.
DA$\Phi$NE, which will produce $K^{\pm}$ and $\klong$
with an annual rate of $9\cdot10^{9}$ and
$1.1\cdot10^{9}$, respectively \cite{dafne},
may improve the upper bounds for $K_{e5}$ decays considerably.

 To summarize, we have evaluated the rates and branching ratios
of $K_{e5}$ decays at leading order in
CHPT (\eqs{gl7}{gl14}) and given the isospin relation
(\ref{gl15}) for the decay modes.
We have furthermore seen that $K_{e5}$ decays,
in particular any effects from chiral anomaly, will be  invisible
at DA$\Phi$NE. However, the upper bounds for
the branching ratios can be improved significantly.

\vspace{0.75cm}
\noindent
I  thank J\"urg Gasser for  useful discussions
and for a critical reading of the manuscript.


\end{document}